\documentclass[review]{elsarticle}

\usepackage{graphicx}
\usepackage{color}
\usepackage{natbib}

\begin{document}

\begin{frontmatter}

\title{Molecular Dynamics Investigation of a Model Ionic Liquid Lubricant for Automotive Applications}

\author[TME]{Konstantinos Gkagkas\corref{mycorrespondingauthor}}
\cortext[mycorrespondingauthor]{Corresponding author}
\ead{Konstantinos.Gkagkas@toyota-europe.com}

\author[Abylsen]{Veerapandian Ponnuchamy}

\author[SCL]{Miljan Da\v{s}i\'{c}}

\author[SCL]{Igor Stankovi\'{c}}

\address[TME]{Advanced Technology Division, Toyota Motor Europe NV/SA, Technical Center, Hoge Wei 33B, 1930 Zaventem, Belgium}
\address[Abylsen]{Abylsen Belgium, 1000 Brussels, Belgium}
\address[SCL]{Scientific Computing Laboratory, Center for the Study of Complex Systems, Institute of Physics Belgrade, University of Belgrade, 11080 Belgrade, Serbia}

\begin{abstract}
In the current work we present a new modelling approach for simulating meso--scopic phenomena related to lubrication of the piston ring--cylinder liner contact. Our geometry allows a variable confinement gap and a varying amount of lubricant in the gap, while avoiding the squeeze-out of the lubricant into vacuum. We have implemented a coarse grain molecular dynamics description of an ionic liquid as a lubricant which can expand into lateral reservoirs. The results have revealed two regimes of lubrication, an elasto-hydrodynamic one under low loads and one with low, velocity-independent specific friction, under high loads. The observed steep rise of normal forces at small plate-to-plate distances is an interesting behaviour that could potentially be exploited for preventing solid--solid contact and wear.

\end{abstract}


\end{frontmatter}

\section{Introduction}

Friction accounts approximately for one-third
of the fuel energy consumed in passenger cars~\cite{holmberg2012carenergy}, therefore a deeper understanding of the lubrication mechanisms in engineering
systems is necessary. Atomic-scale simulations can provide important insights which are necessary for understanding the underlying mechanisms that can affect the
system behaviour, such as structural changes in lubrication layers during shear as well as the interaction between lubricants and surfaces.
The field of computational lubricated nanotribology has been well established over the last decades~\cite{bhushan1995nanotribology, Heo2005}
and the availability of increased computational resources is allowing the application of such methods in cases with increasing complexity. Recent studies
of nanoscopic friction based on Molecular Dynamics (MD) include, for example, the study of fatty acids~\cite{Loehle2014} and ionic
liquids ILs~\cite{mendonca2013ILmetal} as lubricants. Wear reduction demands and the drive to keep friction low, have led to reduced lubricant
film thickness down to only a few molecular layers~\cite{Heyes1.3698601,GattinoniPhysRevE.88.052406,Martinie2016,VoeltzelC5CP03134F}. MD can enable us to access and
understand the behaviour of very thin films which are subjected to compression and shearing between
walls~\cite{Heyes1.3698601,GattinoniPhysRevE.88.052406,Martinie2016}.

Our specific goal is to achieve a representation of the tribological system which is relevant to automotive powertrain applications. As
approximately 45\% of the engine friction losses occur in the piston assembly~\cite{holmberg2012carenergy}, our initial target is to mimic the
conditions observed in the piston ring--cylinder liner contact, in terms of pressure, temperature and shear rates. In addition, in order to be able to
achieve length-- and time--scales that can be of relevance to the real--life systems, it is necessary to apply appropriate simulation methodologies,
such as the use of coarse grain molecular dynamics~\cite{GaoJPCB2004,Robbins2000,PhysRevB.58.R5893,ar700160p}.

In recent years, the application of ILs as advanced lubricants is being studied with a steadily increasing
interest~\cite{Bermudez2009}. The use of ILs as both neat lubricants and additives for engine lubrication has been considered~\cite{Mistry01032009,
am201646k, Qu2009}. Significant improvements on friction and wear reduction have been observed experimentally~\cite{am201646k}, rendering this concept
of potential interest to industry. However, unravelling the mechanism of nanoscopic friction in ILs together with their structure poses a
great scientific challenge, and so far few studies in this direction have been performed, e.g., Ref.~\cite{fajardo2015friction}. ILs are molten salts typically consisting of large-size organic anions and cations. Physical properties of ILs, such as negligible vapour
pressure, high temperature stability (they do not evaporate or decompose at temperatures of interest for automotive industry) and high ionic
conductivity render them potentially relevant to lubrication. In addition, their properties can be modified by an applied voltage using confining
surfaces which are charged in order to build up an electric field across the nanoscale film. The applied potential can affect the structure of
IL layers  and lead to externally controllable lubricating properties~\cite{fajardo2015friction, fajardo2015electrotunable}.
There is also significant flexibility in tuning the physical and chemical properties of ILs which can affect lubrication such
as viscosity, polarity and surface reactivity by varying their atomic composition as well as the anion--cation combination. An important observation
is that ILs confined between surfaces feature alternating positive and negative ionic layers, with an interlayer separation
corresponding to the ion pair size~\cite{lubricants2013,capozza2015squeezout}.

Previous work employing Lennard-Jones fluids has provided insights on the complete dynamic diagram of confined liquids including wall slip, shear banding, solid
friction, and plug flow. In terms of fluid complexity these studies have mainly employed mono-atomic systems, and only a few authors have considered mixtures of molecules~\cite{BarratPhysRevLett.90.095702,BarratPhysRevE.77.021505}. In addition to inherently being a mixture of cation and anion molecules, ILs involve long range Coulomb interactions inducing long range order on far greater scales than the IL itself~\cite{mendonca2013ILmetal,VoeltzelC5CP03134F,CanovaC4CP00005F}. Detailed investigation of ILs as lubricants at the nanoscale is therefore essential for exploring the potential of implementing them in lubrication systems.

In this study, we apply a coarse-grained model for the description of nanoscopic
friction mediated by a liquid lubricant because based on recent studies~\cite{fajardo2015friction, fajardo2015electrotunable,
capozza2015squeezout} it was shown that if the molecules interact
via non-bonded potentials (Lennard-Jones and Coulombic), this can capture all main physical attributes of the IL-lubricated nanotribological system.

This paper is organised as follows: Section 2 introduces the MD setup of the solids and lubricants used, while the motivation for the choices made is provided. In Section 3, the structural properties of the modelled IL under bulk and confined conditions are discussed. The results stemming from the friction MD simulations are then presented in Section 4
followed by some concluding remarks in Section 5.

\section{Model}

Under typical operation of internal combustion engines, the conditions inside the combustion chamber vary significantly. Temperature can range from $300$~K to values higher than $2000$~K, while pressure ranges from atmospheric to values higher than $10$~MPa~\cite{holmberg2012carenergy}. The piston reciprocates with a sinusoidal velocity variation with speeds varying from zero to over $20$~m/s. The time required for one revolution of the engine is of the order of $10$~ms, while the total distance travelled by the piston over this period is of the order of $0.2$~m.
Such scales are typically modelled using continuum mechanics simulations. However such simulations cannot provide the physical insight which is necessary for understanding the molecule--dependent processes that affect the tribological phenomena. For this purpose, we have developed a coarse grain MD configuration that can provide useful insights to molecular processes, while remaining relevant to conditions observed in real--life systems. More specifically, in this section we will describe a
setup of MD simulations developed with the aim of building a framework that incorporates meso--scale features of the piston ring--cylinder liner system and permits an efficient implementation of different solid surfaces and lubricants.

\subsection{Geometry}

\begin{figure*}[htb]
\includegraphics[width=0.8\textwidth,natwidth=1089pt,natheight=751pt]{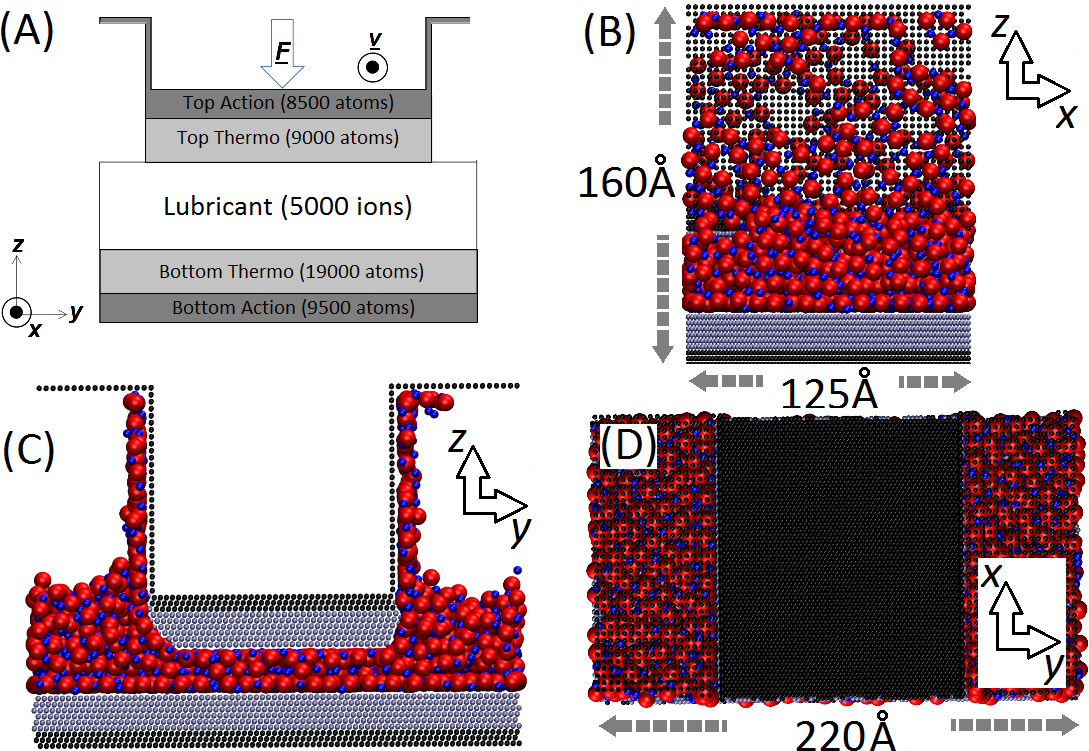}
\caption{(A) Schematic of the simulation setup shown as yz cross-section.
There are two solid plates at the top and bottom of the system, consisting of two regions: at the outermost ones the particles are moving as a single entity (Top/Bottom Action) and at the innermost ones the particles are at a controlled temperature (Top/Bottom Thermo). The thermalised layers are in direct contact with the lubricant while the action layers are used to impose external velocity and/or force to the solid plates. (B)-(D) Side views of the typical simulation configuration and key dimensions of the geometry. (B) Side ($xz$) view with respect to the shear direction. (C)  Front ($yz$) view in the direction of the top plate motion.
(D) Top ($xy$) view of the system. The solid plates are made up of FCC (111) atomic layers.
The ionic liquid is composed of an equal number of cations (blue spheres) and anions (red spheres).
}
\label{fig:geometry}
\end{figure*}

All MD simulations in this study were performed using the LAMMPS software~\cite{plimpton1995fast}.
The geometry used in our friction simulations is shown as a schematic in Figure~\ref{fig:geometry}, along with the dimensions of our system as well as the number of the MD particles used. The simulation setup was loosely inspired by previously published research by others~\cite{mendonca2013ILmetal,fajardo2015friction,fajardo2015electrotunable,capozza2015squeezout}. By implementing such a geometry we have attempted to achieve: $(i)$ a realistic particle squeeze--out behaviour with the formation of two lateral lubricant regions (in a similar manner to the simulations of
Capozza et al.~\cite{capozza2015squeezout}) and $(ii)$ a system that allows the lubricant to be externally pressurised. For the description
of the solid surfaces we have combined rigid layers of particles moving as a single entity
on which the external force or motion is imposed, denoted by "Top Action" and "Bottom Action" in Figure~\ref{fig:geometry}(A),
with thermalised layers (denoted by "Top Thermo" and "Bottom Thermo") in which particles vibrate thermally at $T=330$~K. The Nose-Hoover NVT thermostat is used to control the temperature. As in previous MD simulations~\cite{VoeltzelC5CP03134F,
fajardo2015friction,fajardo2015electrotunable,capozza2015squeezout,CanovaC4CP00005F}, under similar operating conditions, the details of the adopted dissipation scheme are not expected to change the essence of the system response on mechanical deformation.  The relaxation time of the Nose-Hoover NVT thermostat for the lubricant and the solids is $200$~fs (cf. Ref.~\cite{VoeltzelC5CP03134F}). The plates were treated as rigid bodies, with the
lower one being fixed and the upper one subjected to a z-directed force $F_z$ (the load) as shown in Fig.~\ref{fig:geometry}(A) and
driven along $x$ direction at constant velocity. The solid plates were made up of densely packed atomic layers at a FCC (111) lattice arrangement. Periodic
boundary conditions were applied in the $x$ and $y$ directions. The bottom plate can therefore be considered to be infinite, while the top plate is surrounded by vacuum pockets on its sides, so called lateral reservoirs, in which the lubricant can freely expand. The lateral reservoirs were implemented as a mechanistic way for allowing the lubricant to be dynamically squeezed in or out as an external load or velocity is applied, or due to local fluctuations during the simulation progression. At the same time, the lubricant remains an infinite continuous body in $x$ and $y$ directions, similar to the conditions observed in a real tribological system from a meso--scopic point of view. This is especially important if the system experiences partial or complete crystallisation under compression, cf. Section 4 and Fig.~\ref{fig:cryst_snaps}.

While the total number of considered lubricant molecules is constant, the  finite upper plate width and the resulting free space enables the
particles to be squeezed-out even to the extent where due to structural changes the lubricant ceases wetting the solid plates. The
number  of  lubricant molecules  effectively  confined  inside  the  gap can therefore dynamically change depending
on the loading conditions. This is important for exploring the possible states of a mechanical system of particles
that is being maintained in thermodynamic equilibrium (thermal and chemical) with a lubricant reservoir (i.e., void spaces in
tribological system). The nano-tribological system is open in the sense that it can exchange energy and particles,
realising an effectively grand-canonical situation, cf. Fig.~\ref{fig:geometry}(b) and Ref.~\cite{GaoPRL1997}.

\subsection{Solids and lubricant model}

By using our simulation setup, we aim to study the lubrication properties of several lubricants. As a first step, in the current study we have implemented an ionic liquid as a lubricant. On the atomic level ILs are usually made up of large-size irregular
organic anions and cations often including long alkyl chains. In the current work we have applied a simple coarse-grained model
for its description, consisting of a charged Lennard-Jones system where anions and cations have different radii as already exploited
in previous studies in the literature~\cite{capozza2015squeezout}. According to that,
we have implemented a Lennard-Jones (LJ) 12-6 potential combined
with a Coulombic electrostatic potential:

\begin{equation}
V\left(r_\mathrm{\textit{ij}}\right) = 4 \epsilon_\mathrm{\textit{ij}}
\left[\left(\frac{\sigma_\mathrm{\textit{ij}}}{r_\mathrm{\textit{ij}}}\right)^{12} -
\left(\frac{\sigma_\mathrm{\textit{ij}}}{r_\mathrm{\textit{ij}}}\right)^6\right] + \frac{1}{4 \pi \epsilon_0 \epsilon_r} \frac{q_i
q_j}{r_\mathrm{\textit{ij}}}
\end{equation}

Parameters $\left\{\epsilon_\mathrm{\textit{ij}}, \sigma_\mathrm{\textit{ij}}\right\}$ define the LJ potential between different types of particles $i,j={{\rm A},{\rm C}, {\rm P}}$ which refer to anions, cations and solid plate atoms, respectively. The numerical values for each pair are listed in Table~\ref{tab:tab1}. The diameter of cations was set to $\sigma_{\rm CC} = 5$~{\AA} and anions to $\sigma_{\rm AA} = 10$~{\AA}, in order to explore the effect of asymmetry of ion sizes (similar to Ref.~\cite{capozza2015squeezout}). Atoms of the solid plates have a diameter of $\sigma_{\rm PP} = 3$~{\AA}.
The plates consist of nine densely packed layers in a FCC ${\rm (111)}$ lattice.

The ions were modelled as coarse grain particles, the charge of which was set to elementary:
$q_{\rm A}=-e$ and  $q_{\rm C} =+e$, i.e., $e= 1.6 \times 10^{-19}$~C. The ionic liquid is neutral, so the total number of cations and anions is the same: $N_{\rm C} = N_{\rm A} = N_\mathrm{IL}/2$. In the present simulations, the number of ions used was $N_{\rm IL} = 2500$. The dielectric constant was set to $\epsilon_r = 2$ to account for the dielectric screening, as in Refs.~\cite{fajardo2015electrotunable} and~\cite{capozza2015squeezout}. The LJ potential has a short-range impact, since it vanishes rapidly as the distance increases $\propto r^{-6}$, while the Coulombic potential has a
long-range impact, $\propto 1/r$. To handle long--range interactions, we have used a multi-level summation
method (MSM)~\cite{hardy2009multilevel}, since it scales well with the number of ions and allows the use of mixed periodic and non-periodic boundaries that are featured in our setup. To sum up, IL ions and plate atoms interact with each other via LJ potentials.
In addition a Coulombic electrostatic potential is added in ion-ion interactions.

In engineering applications, the lubricant molecules typically interact with metal surfaces.
Computationally efficient many--body potentials such as embedded atom method (EAM) potential
~\cite{daw1984embedded,stankovic2004structural}
can be applied for the description of such solids.
Such schemes have been employed extensively for modelling a wide range of systems including
metals ~\cite{stankovic2004structural}
and metal-metal oxide interfaces~\cite{gubbels2014ionic}. In addition, in order to account for the induced charges on the metallic conductor surface by the ions, the Drude-rod model developed by Iori and Corni~\cite{JCC:JCC20928} which consists of the addition of an embedded dipole into each metal atom, thus rendering it polarisable, has been applied in previous studies~\cite{mendonca2013ILmetal}. Since in our initial stage
of IL tribological behaviour research,
modelling the elasticity of metallic plates plays a secondary role, we have selected a simplified model in which plate
atoms interact strongly with each other if they belong to the same plate, i.e., $\epsilon_{\rm PP}=120$ kCal/mol,
as opposed, to a very weak interaction between the different plates $\epsilon_{\rm top/bottom}=0.03$ kCal/mol. Furthermore, even though the typical engineering systems are often metallic, our initial coarse grain MD study of liquid dynamics according to the applied conditions justified the implementation of a simpler solid system which does not feature substrate polarisation. Finally, it is possible that the actual surfaces might feature carbon coatings or depositions, in which case the surface polarisation can be of secondary importance.

\begin{table}
\begin{center}
\begin{tabular}{ |c|c|c| }
 \hline
 pair $ij$& $\epsilon_{ij}$~[kCal/mol] & $\sigma_{ij}$~[{\AA}] \\
 \hline
 \hline
 CC & 0.03 & 5 \\
 \hline
 AA & 0.03 & 10 \\
 \hline
 CA & 0.03 & 7.5 \\
 \hline
 PC & 0.3 & 4 \\
 \hline
 PA & 0.3 & 6.5 \\
  \hline
 PP & 120 & 3 \\
 \hline
\end{tabular}
\end{center}
\caption{List of LJ parameters used in simulations.}
\label{tab:tab1}
\end{table}

The starting configuration for our MD simulations was obtained via a relaxation process. In order to obtain a stable and reproducible initial configuration, the relaxation was performed through a stepwise increase of absolute ion charge at steps of $\Delta|q_i|=e/10$, $i=\rm{A, C}$. Each time the charge of the ions was increased, a
minimisation of the system's energy through conjugated gradient method was performed. In this way, the system characteristics were gradually converted from pure LJ to a Coulomb interaction dominated system through a series of stable configurations.

As we are aiming at understanding the lubricant behaviour at meso-scopic conditions observed in a ring--liner system, we have attempted to include in our MD model the potential IL pressurisation that can occur due to the liquid flow resistance, as well as the variable pressure arising from the reacting gas in the combustion chamber. Inserting gas molecules directly
in the
simulation for this purpose would require a reduction of the time step due to higher thermal velocities of the gas. In turn, the computational cost would increase
significantly making simulations impossible to run in realistic computational time. Therefore, in order to understand the effect of external pressure on the IL behaviour, we have applied a repulsive force between the topmost rigid solid layer and the IL particles in the form of a truncated and shifted LJ potential. Two cases with cut-off distances at $15$~{\AA} and $4$~{\AA} above the outermost top plate layer were studied so that the IL inside the confinement gap would remain outside the influence zone of this mechanistic force.
By appropriate selection of the LJ parameters for this potential, the resulting external pressures applied on the unconfined surface of the IL were $120$~kPa and $250$~kPa, respectively.

\section {Probing ionic liquid's internal structure behaviour}

\subsection{Solidification and melting of bulk ionic liquid}

In order to confirm that the IL used in our MD simulations remains in a liquid state for the conditions of interest, its liquid--solid and
solid--liquid phase transitions were initially studied. A bulk IL configuration was prepared by placing the same number of cations and anions $N_c =
N_a = 1000$ into a 3D periodic box, with pressure kept constant at $100$~kPa. Phase transitions were then achieved via the application of linear
ramping to the temperature, in a similar approach to the calculations performed in Ref.~\cite{capozza2015squeezout}.

\begin{figure*}[htb]
\includegraphics[width=0.8\textwidth,natwidth=2400pt,natheight=1485pt]{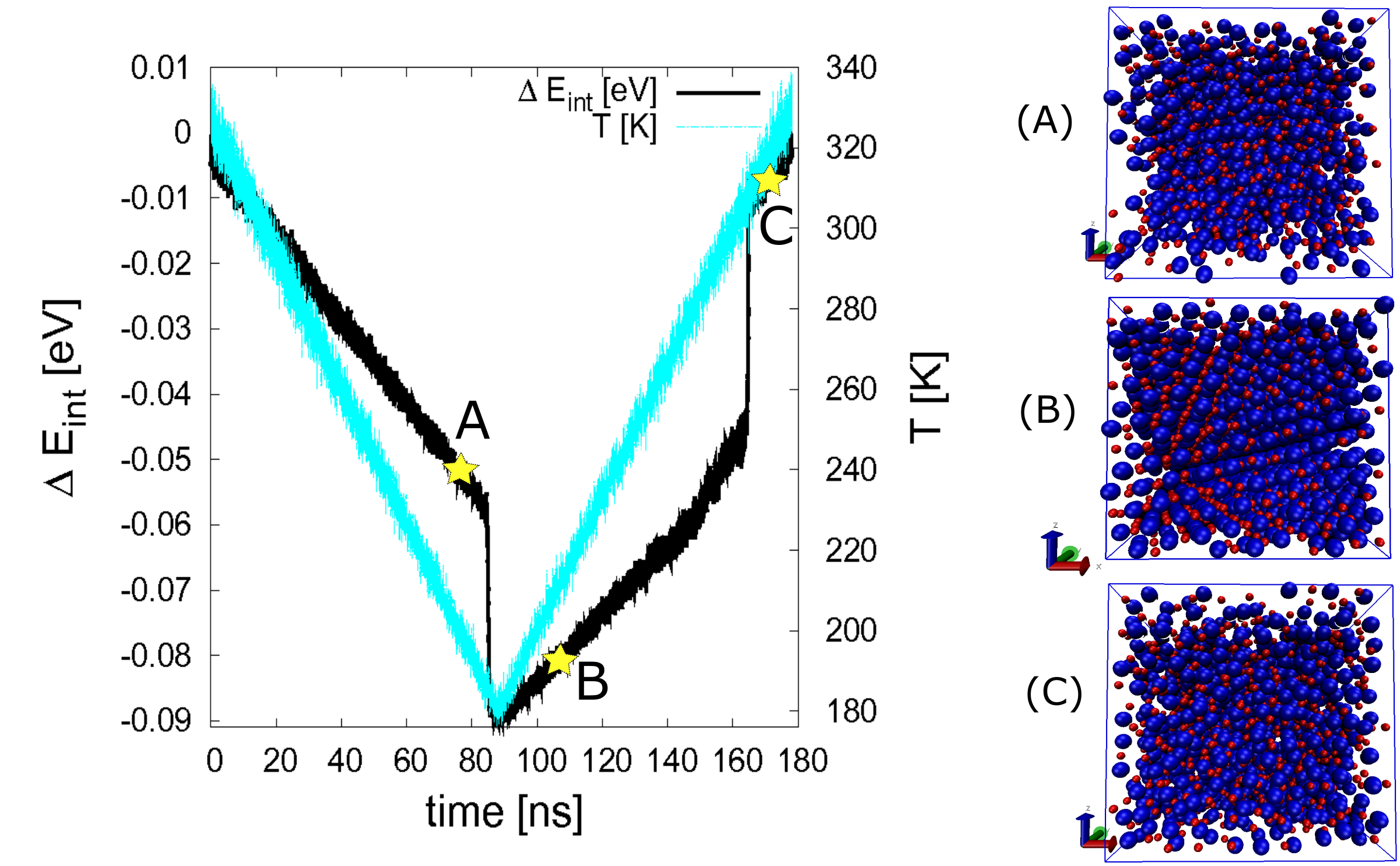}
\caption{(Left): Bulk internal energy change and temperature of the ionic liquid as a function of
simulation time. (Right) Snapshots of ion arrangement at liquid (A), (C) and solid (B) state.
}
\label{fig:ETvstime}
\end{figure*}

\begin{figure*}[htb]
\includegraphics[width=0.8\textwidth,natwidth=1630pt,natheight=1364pt]{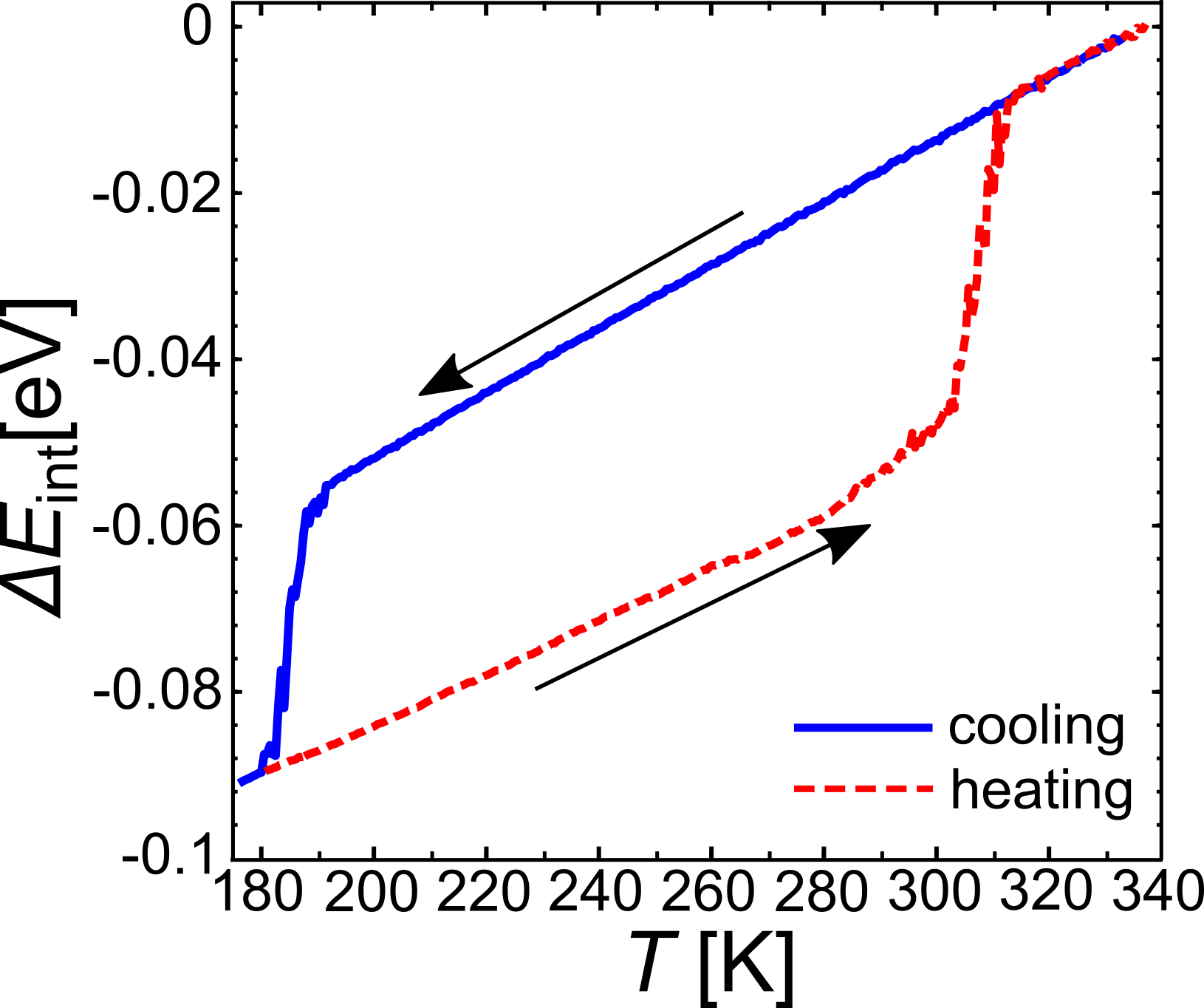}
\caption{Bulk internal energy change of the ionic liquid as a function of temperature. The internal energy was calculated by averaging on segments of $\Delta T = 0.5 K$.
}
\label{fig:Eint_T}
\end{figure*}

Starting from an initial temperature $T_1 = 330$~K where the IL is in liquid state, the temperature was decreased linearly down to $T_2 = 180$~K.
The absolute rate of temperature change was: $\left|dT\right|/dt = 1.67$~$K$~$ns^{-1}$. A liquid--solid phase transition was observed during the IL cooling. After reaching $T_2 = 180$~K, the temperature was increased back to the initial value of $330$~K. This heating process caused with its turn a solid--liquid phase transition. In Fig.~\ref{fig:ETvstime} the IL internal energy change $\Delta E_\mathrm{int}$
and temperature $T$ are shown as functions of time $t$. The temperature profile follows the applied conditions and its superimposition to internal
energy change allows the observation of the dynamic behaviour of the liquid. By plotting the averaged internal energy change of the IL against its temperature in Fig.~\ref{fig:Eint_T}, the hysteresis behaviour in the solidification--melting cycle is clearly observed, while the phase transition locations can be clearly defined. It can be seen that during the cooling process, the internal energy of IL linearly decreases until the temperature reaches $T_\mathrm{ls} ~= 190$~K, at which point a sharp drop is observed. This indicates a first order thermal phase transition (liquid--solid). During the heating process, a similar sharp jump of energy is observed at $T_\mathrm{sl} ~= 305$~K which corresponds to an opposite phase transition (solid--liquid).  The obtained results are in a good agreement with Ref.~\cite{capozza2015squeezout} and confirm that the IL is behaving as a liquid for temperatures higher than $310$~K, under atmospheric pressure conditions. In the rest of our calculations a temperature value of $T = 330$~K was applied, in order to allow a liquid state that is combined with local
solidification under elevated contact pressure conditions.

\subsection{Ionic liquid structure in thin film}

\begin{figure}[htb]
\includegraphics[width=0.8\textwidth, natwidth=1901pt,natheight=1138pt]{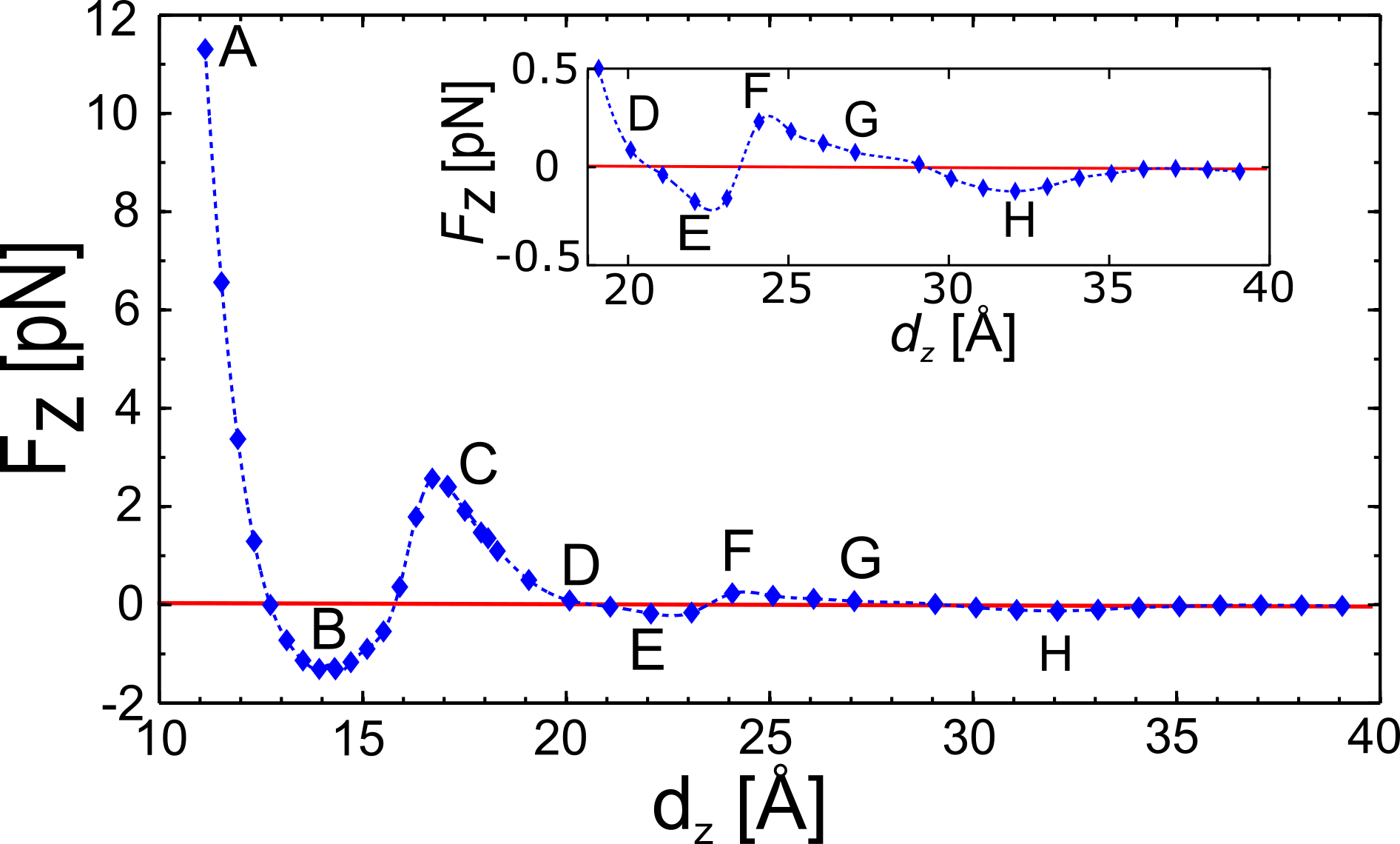}
\caption{Dependence of normal force $F_{z}$ on plate-to-plate distance $d_{z}$. Eight characteristic points $\left\{A, B, C, D,
E, F, G, H\right\}$
with corresponding interplate distances $d_z = \left\{11, 14, 17, 20, 22, 24, 27, 32\right\}$~{\AA} are marked on the $F_z\left(d_z\right)$ curve. The horizontal
solid line denotes $F_z = 0$~pN. The dashed line connects the points obtained from the simulation and serves as a visual guide.
}
\label{fig:FzvsDz}
\end{figure}

The confinement has a profound influence on the structure of ILs in thin films~\cite{lubricants2013,GaoPRL1997,PerkinPCCP2012}. The confining surfaces can induce ordering of the particles in their vicinity. The resulting structure and forces are a result of the interplay between the limited volume and the particles which fill the space. In  Fig.~\ref{fig:FzvsDz}, the force-distance characteristic of our system is presented. The red horizontal line denotes the zero normal force level (i.e., $F_z=0$). A non-monotonous behaviour of the normal force $F_z$ acting on the top plate can be observed as the plate-to-plate distance is changing. This distance corresponds to the gap between the plates where the IL is under confinement. The points $\left(d_z, F_z\right)$ were obtained through our simulations, while the dashed line serves as a visual guide. It can be seen that the normal force strongly depends on the inter--plate distance and that it also becomes negative in certain regions. This can be translated as the IL striving to reduce the plate-to-plate distance due to adhesion phenomena. These changes of the normal force are correlated with the extraction and inclusion of IL layers into the gap, as already observed experimentally, cf. Ref.~\cite{lubricants2013}. During the performed simulations, the cationic-anionic layers were squeezed out in pairs, in order to keep the system locally neutral, as observed in experimental studies~\cite{lubricants2013, GaoPRL1997, PerkinPCCP2012, hayes2011double, smith2013quantized}.

Concerning the realisation of the simulations presented in Fig.~\ref{fig:FzvsDz}, the inter--plate gap was modified in the following manner: the top plate was displaced towards the bottom one with a constant velocity $v_z = 5$~m/s. For $d_z<17$~{\AA} the velocity was reduced to $v_z = 1$~m/s. At each calculation point shown in Fig.~\ref{fig:FzvsDz},
the top plate was kept fixed for a period of time $t_{static} = 50$~ps, during which period the average value of the normal force was calculated. The process was repeated until a distance $d_{z,min} = 11$~{\AA} was reached.

\begin{figure}[htb]
\includegraphics[width=0.8\textwidth, natwidth=1626pt,natheight=1580pt]{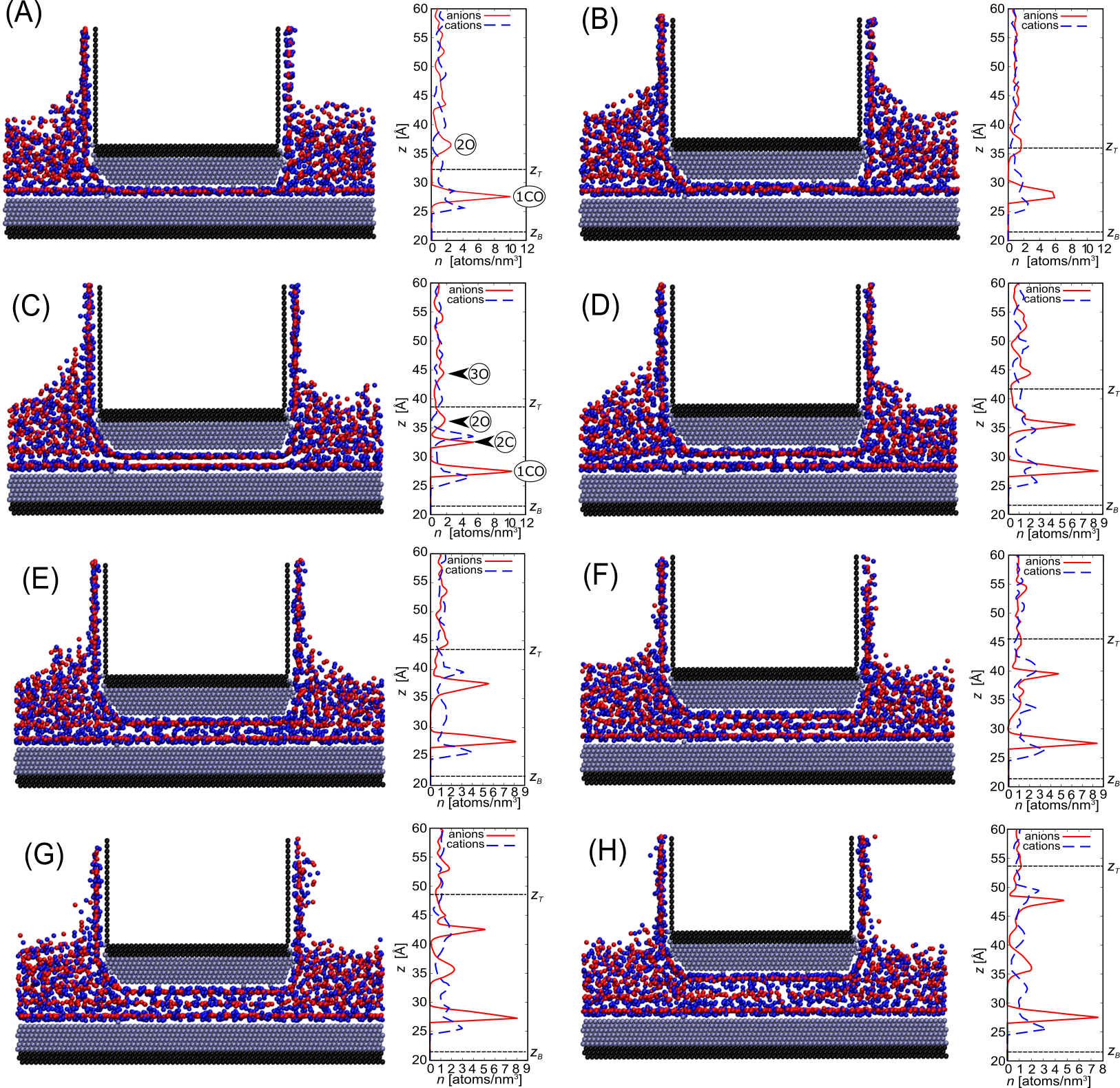}
\caption{Snapshots of system configurations at points $\left\{A, B, C, D, E, F, G, H\right\}$ from Fig.~\ref{fig:FzvsDz} and
corresponding density distribution of anions/cations along the $z$-axis. The position of the atomic centres of the innermost layer
of the top and bottom plate is denoted by $z_{\rm T}$ and $z_{\rm B}$, respectively. The bottom plate is fixed and $z_{\rm B}=21$~{\AA}. The ions are
deliberately depicted smaller than their LJ radii in order to allow a direct observation of the layering. In Figures (A) and (C) the annotations indicate the anion layer vertical order from the bottom (1, 2, 3) and the lateral placement: (C)onfined and (U)nconfined.
}
\label{fig:Dz_snapshots}
\end{figure}

In order to understand the dynamic evolution of our system, snapshots of the system from the MD simulations corresponding to several characteristic points
in the $F_z \left(d_z\right)$ curve were selected and studied in more detail. Fig.~\ref{fig:Dz_snapshots} shows  the configuration and ion density distribution along the z--direction for eight characteristic points $\left\{A, B, C, D, E, F, G, H\right\}$,
corresponding to plate-to-plate distances $d_z = \left\{11, 14, 17, 20, 22, 24, 27, 32\right\}$ {\AA} respectively.
The ions are deliberately depicted smaller than their LJ radii in order to allow a direct observation of the layering. The position of the atomic centres of the innermost atomic layers of the top and bottom plate are indicated in Fig.~\ref{fig:Dz_snapshots} as $z_{\rm T}$ and $z_{\rm B}$ respectively. As the bottom plate was fixed, $z_{\rm B}$ remains constant while $z_{\rm T}$ changes with the top plate displacement.

A general feature observed under all conditions was the fact that the cations always formed the layer closest to the bottom plate.
The reason is the smaller size of the cations ($\sigma_{CC}=5$~{\AA}) compared to the anion species ($\sigma_{AA}=10$~{\AA}).
Following this, a second layer was induced by the first one and populated only by anions.
The distance between the first and the second layer from the bottom is in the range of $1-2.5$~{\AA}, meaning that while the centres
of mass of the particles are in different layers, the layers themselves overlap as their distance is smaller than the particle diameters.
In the rest of this section, the changes in the number of layers as the inter--plate gap is reduced will be presented and correlated with the changes in the normal force $F_z$ which is acting on the top plate.

For the minimum simulated plate-to-plate distance $d_z = 11$~{\AA}, shown in Fig.~\ref{fig:Dz_snapshots}(A) we can observe a pronounced peak in the anion density distribution close to the bottom plate which is aligned with a well-defined anionic layer inside the gap. The anion peak is marked with the ``1CU'' annotation. In the case of cations, there are two peaks attached below and above the anionic peak. This situation corresponds to the formation of two incomplete cationic layers inside the gap. With increasing plate-to-plate distance $d_z$ the normal force $F_z$ is decreasing, with a sign change of $F_z$ at $d_z = 12.7$~{\AA}. In the range $12.7$~{\AA}$ < d_z < 15.7$~{\AA} the normal force remains negative. This means that the IL is pulling the plates together, since the ions strive to reduce their interlayer distance, as well as the distance between themselves and the plate atoms. Such behaviour is typically observed in systems exhibiting layering transition, already seen in systems of both neutral molecules ~\cite{bhushan1995nanotribology} and ILs~\cite{lubricants2013}. With further increase of $d_z$ the force becomes positive again, and reaches a local maximum at the point (C) in Fig.~\ref{fig:FzvsDz}. At this point we observe a change in the number of anion layers confined in the gap from one to two, as shown in Fig.~\ref{fig:Dz_snapshots}(C).

In Fig.~\ref{fig:Dz_snapshots}(C), the plate-to-plate distance is $d_z=17$~{\AA} and the two bottom peaks of the anion/cation density distribution, denoted by ``1CU'' and ``2C'', are inside the gap. A third smaller anion/cation density peak, denoted by ``2U'' in Fig.~\ref{fig:Dz_snapshots}(C), is the result of the ordering initiated at the bottom plate's surface and is actually outside the confinement gap. The vertical distance between the peaks ``2C'' and ``2U'' is approximately $3.5$~{\AA} and corresponds to the effect of the compression of the IL from the top plate. Further increase of the plate-to-plate distance results in a continuous decrease of the normal force without a sign change as the positions of peaks ``2C'' and ``2U'' become aligned, cf. Fig.~\ref{fig:Dz_snapshots}(D) for a distance $d_z=20$~{\AA}. Further increase of the inter--plate distance results once more in a reversal of the sign of the normal force (i.e., $F_z<0$ for $21$~{\AA}~$<d_z<23.5$~{\AA}). At the mid point between the plates a broad maximum of cation density distribution can then be observed, see Fig.~\ref{fig:Dz_snapshots}(E). The cations, as smaller particles, have a tendency to fill the space between the more stable anionic layers. When the anions also start to form a third layer at the midpoint between the two plates the corresponding cationic peak of density becomes sharper and the normal force becomes positive again, see Fig.~\ref{fig:Dz_snapshots}(F). In this case the cations can from a layer more easily while the anions remain scattered. This is the opposite behaviour to the one typically observed, where the larger anions tend to order more strongly due to the excluded volume effect~\cite{kaplan1994entropically}. From Fig.~\ref{fig:Dz_snapshots}(F) to Fig.~\ref{fig:Dz_snapshots}(G) an interesting transition can be observed, during which the single well resolved cation peak disappears and a less pronounced cation--anion pair peak takes its place. Finally in Fig.~\ref{fig:Dz_snapshots}(H) at $d_z = 32$~{\AA}, we observe the clear formation of three anion and four cation peaks.

Considering engineering applications, the steep rise of the normal force at small plate-to-plate distances, i.e., $d_z < 14$~{\AA} can be beneficial for protecting against solid-solid contact and consequent wear. On the other hand, there is also a strongly decreasing trend of maximal normal force
which can be sustained by the system as the number of ion layers confined between the plates increases, i.e., for two cation layers the maximal force $F_{z,max}=3$~pN, while for three it is $F_{z,max}=0.25$~pN. In our model, the Lennard-Jones interaction between the plates and the ions is ten times stronger than between the ions themselves. The ion layers closest to the plates are therefore more stable than the layers in the midpoint of the gap (see Fig.~\ref{fig:Dz_snapshots}(F)). As a result, the three layer system becomes less dense, and can build up a lower normal force compared to the two layer system (in Fig.~\ref{fig:Dz_snapshots}(C)).

\section{Friction simulations}

\begin{table}
\begin{center}
\begin{tabular}{ |l|c|c|c| }
 \hline
 Case & $a$ & $b$ & $R^2$ \\
 \hline
 \hline
(A) $d_z=17$~{\AA}, $p_{\rm ext}=0$~kPa & -0.0006(2) & 0.0039(2) & 0.63\\
 \hline
(B) $d_z=27$~{\AA}, $p_{\rm ext}=0$~kPa & 0.016(5) & 0.036(3) & 0.72 \\
 \hline
(C) $d_z=27$~{\AA}, $p_{\rm ext}=120$~kPa & 0.007(2) & 0.017(2) & 0.26 \\
 \hline
(D) $d_z=27$~{\AA}, $p_{\rm ext}=250$~kPa & 0.002(1) & 0.003(1) & 0.62 \\
 \hline
\end{tabular}
\end{center}
\caption{Results for the coefficients $a, b$ in the relation $\langle F_x\rangle/\langle F_z\rangle=a\log(v_x/v_{\rm ref})+b$, where $v_{\rm ref}=1$~m/s. The results were obtained using the least-squares method.}
\label{tab:tab2}
\end{table}

\begin{figure*}[htb]
\includegraphics[width=0.8\textwidth,natwidth=2100pt,natheight=1600pt]{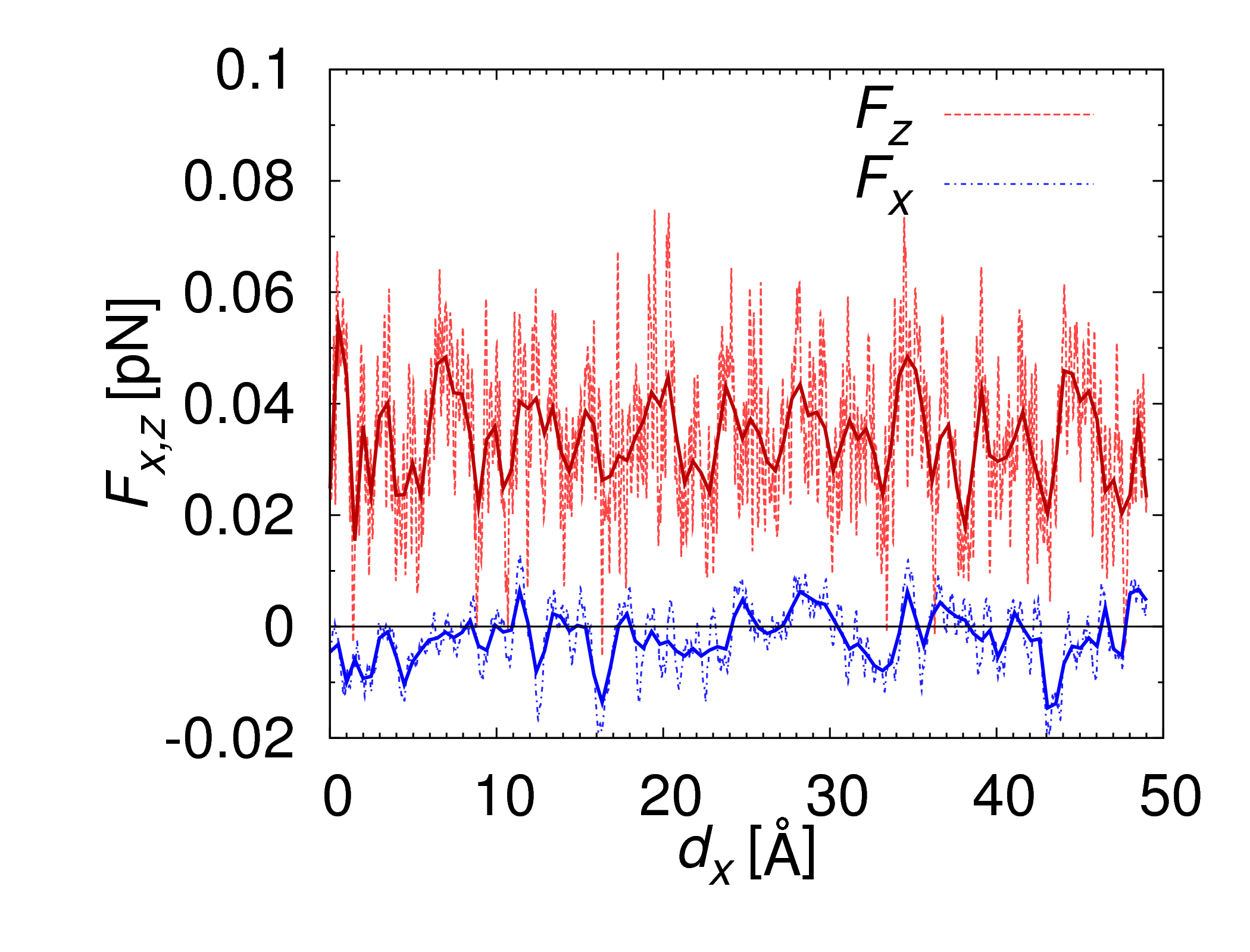}
\caption{Temporal evolution of total normal and axial forces acting on sliding surface for plate-to-plate distance $d_z=27$~{\AA} and top plate axial velocity $v_x=10$~m/s. Dashed lines show the raw numerical data which are smoothed using the solid lines for a clearer identification of temporal trends.
}
\label{fig:force_time}
\end{figure*}

Following the detailed study of the static system, we turn our focus to dynamic conditions, where there is a relative motion between the plates in $x$ direction and as a result frictional forces can be observed. The dynamics of the plates impact the IL and result in an overall longitudinal force acting on each solid body. In order to evaluate the trends of specific friction we have performed simulations at different plate velocities and at two interplate distances.  The simulations have been performed for a broad range of top plate velocities $v_x=0.1,$ $0.2,$ $0.5,$ $1,$ $2,$ $5,$ and $10$~m/s, with the bottom plate kept fixed. We have compared cases with different external pressures applied on the IL $p_{\rm ext}=0,$ $120$  and $250$~kPa and two distinct plate distances $d_z=17$ and $27$~{\AA}.The simulations were performed as follows: Points (C) and (G) in Fig.~\ref{fig:Dz_snapshots} were chosen as the starting configurations. The simulations ran until the top plate had covered a distance of $d_x=50$~{\AA} in $x$ direction. Therefore cases with lower velocities required increased total time. The forces acting on the top plate were monitored, as shown in Fig.~\ref{fig:force_time} for a randomly chosen case. It was observed that the normal force remained roughly the same after the onset of the simulation. Steady--state conditions were assumed following a displacement of $d_x=10$~{\AA}, and then average values were calculated using the statistics until the completion of the simulation.

\begin{figure}[htb]
\includegraphics[width=0.8\textwidth,natwidth=1718pt,natheight=1353pt]{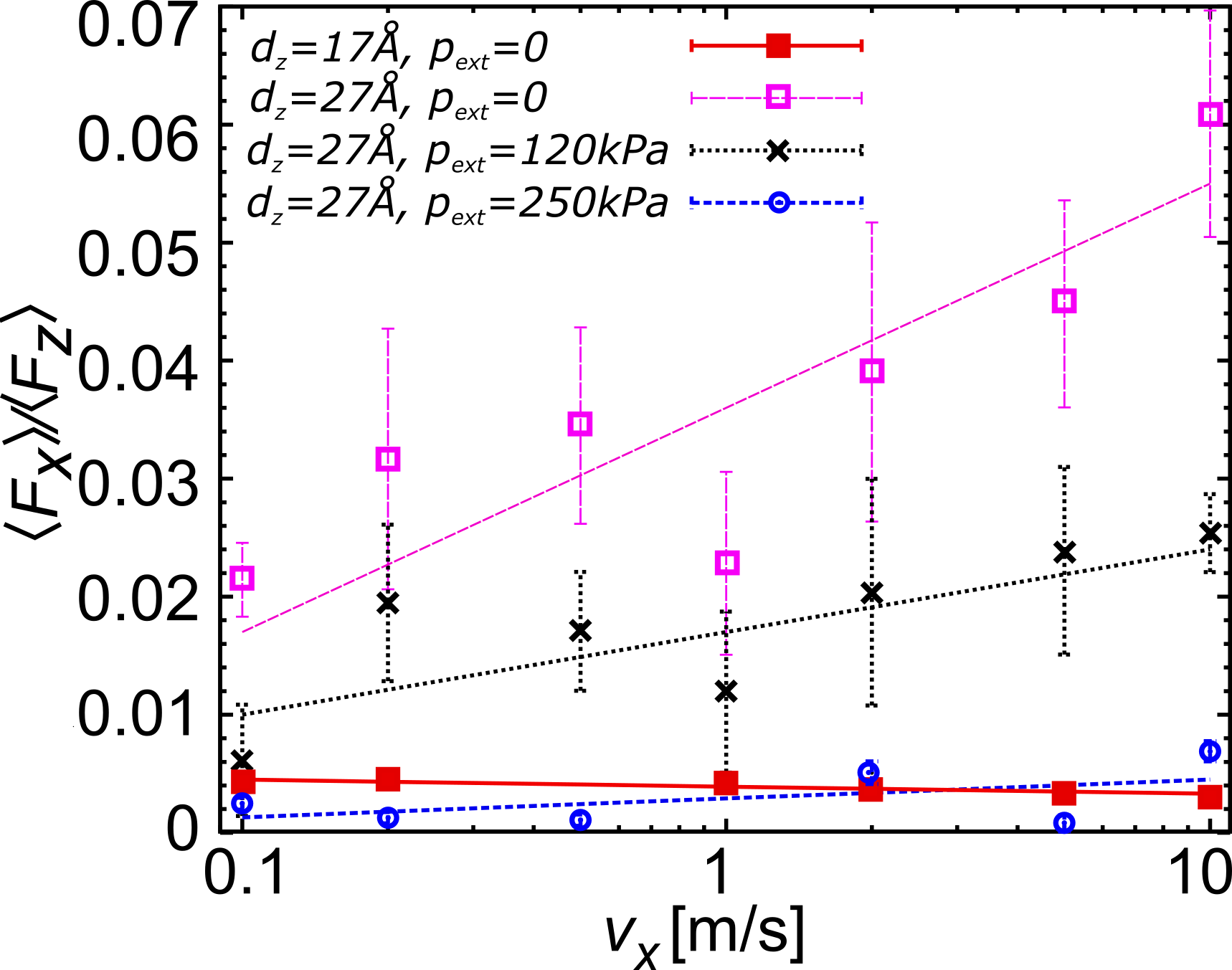}
\caption{Dependence of specific friction $\langle F_x\rangle/\langle F_z\rangle$ on velocity at external pressures $p_{\rm
ext}=0,$ $120$ and $250$~kPa and inter-plate distances $d_z=17$ and $27$~{\AA}. The error bars represent the standard deviation of the average values obtained from the simulation data. The curves showing the specific friction trends were obtained by linear regression and the corresponding parameters are listed in
Tab.~\ref{tab:tab2}
}
\label{fig:cof_Vx}
\end{figure}

The results for the specific friction are shown as a function of sliding velocity in Fig.~\ref{fig:cof_Vx}. The specific friction $\langle F_x\rangle/\langle F_z\rangle$ is defined as the ratio of the time averaged frictional and normal force $F_x$ and $F_z$ respectively and is different to the Coulomb friction coefficient $\mu=\partial{F_x} / \partial{F_z}$. In our simulated cases we have observed either a weak or a logarithmic dependence of specific friction on velocity. The numerical values were fitted to a linear function of the form $\langle F_x\rangle/\langle F_z\rangle=a\log(v_x/v_{\rm ref})+b$, where $v_{\rm ref}=1$~m/s. The coefficients $a,b$ obtained from the simulation data are listed in Table~\ref{tab:tab2}. A reasonable fit to the linear regression curve can be observed for most cases. In the case of $p_{\rm ext}=120$~kPa, the system is potentially in a transition between the two significantly different cases of zero and high pressure, which can explain the poorer quality of the fit to the linear curve. The logarithmic dependence indicates typical elasto-hydrodynamic lubrication conditions~\cite{Bair2016}. On the other hand, the weak dependence of specific friction on velocity has also been observed in previous studies of IL lubrication, cf. Ref.~\cite{mendonca2013ILmetal,CanovaC4CP00005F}.

\subsection{Impact of ionic liquid confinement gap}

The influence of plate-to-plate distance on specific friction was initially analysed, while the applied external pressure on the IL $p_{\rm ext}$ was kept equal to zero. In contrast to previous studies of IL lubrication~\cite{mendonca2013ILmetal, CanovaC4CP00005F}, our system showed a strong crystalline order induced by confinement. The normal force was
roughly ten times higher in the case of the smaller plate-to-plate distance, i.e., for $d_z=17$~{\AA} compared to $d_z=27$~{\AA}. On the other hand, the lateral force $F_x$ remained at similar levels, therefore leading to a sharp
decrease of the specific friction values. At the same time, the weaker confinement and the smaller normal force for $d_z=27$~{\AA} resulted in a steeper slope of the curve $\langle F_x\rangle/\langle F_z\rangle$.

\begin{figure}[htb]
\includegraphics[width=0.8\textwidth,natwidth=2127pt,natheight=2618pt]{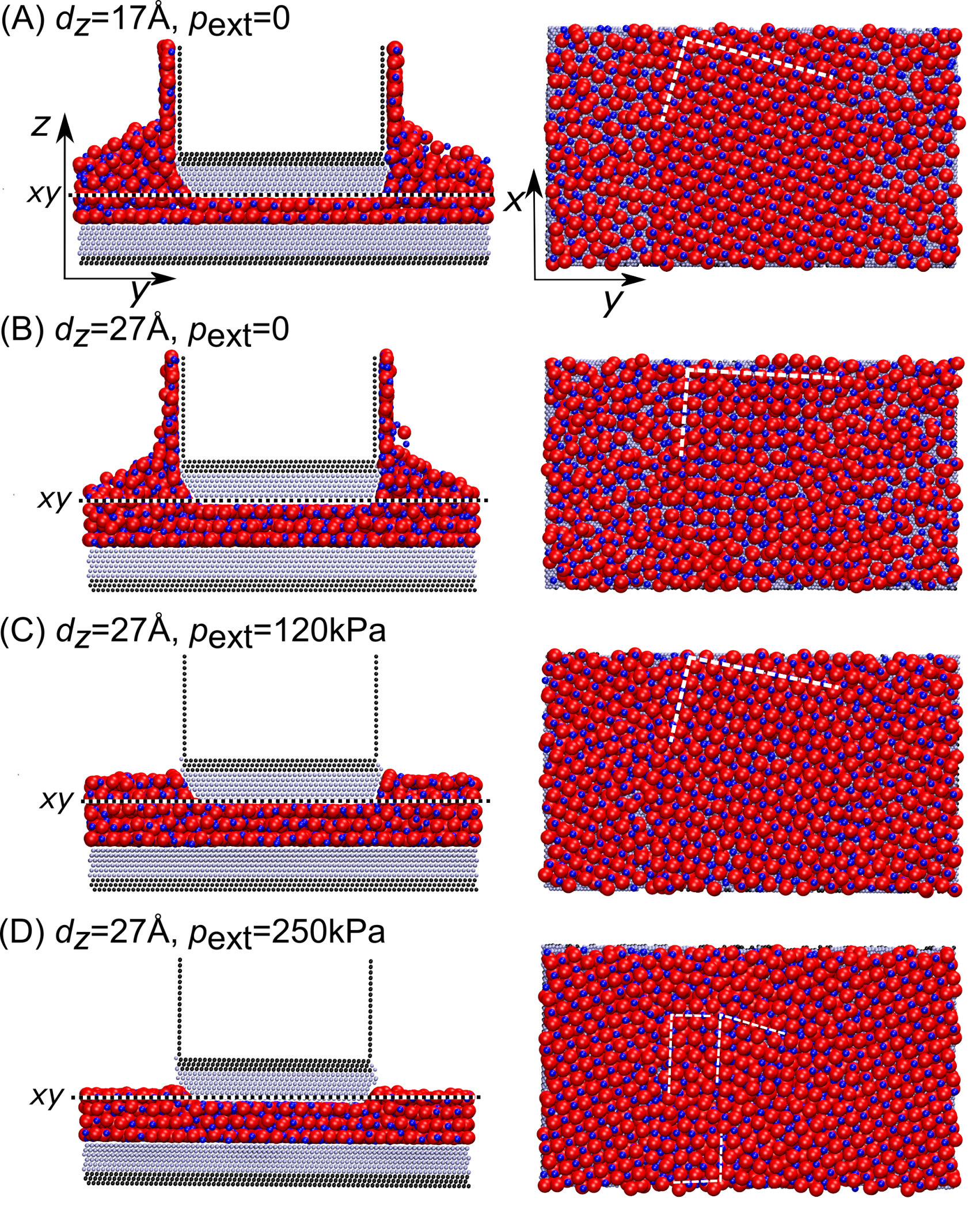}
\caption{Side ($yz$) and top ($xy$) views of snapshots from four separate friction simulations. The top views correspond to the planes marked with dashed lines in the side views and do not include the solid and IL particles above the upper plate plane. The ions are depicted according to their LJ radii in order to visualise the crystalline structures. The dashed lines in the top views denote the crystal
direction of self-formed cubic structures.}
\label{fig:cryst_snaps}
\end{figure}

In order to understand the potential correlation of the IL structure with the arising frictional forces, the confinement zone was observed in detail using Fig.~\ref{fig:cryst_snaps}, where a side view (left side) and top view (right side) of the system is shown.  In the top view, the system is shown with the solid and IL particles above the upper plate plane removed. In this plot the ions are depicted with their corresponding LJ radii in order to achieve a realistic visualisation of the structure. The anions form a locally cubic structure, cf. right panel Fig.~\ref{fig:cryst_snaps}(A), while the crystal direction of the cubic structure is indicated with dashed lines. If we look into the structure of the IL in the confinement zone, Fig.~\ref{fig:cryst_snaps}(A) and (B), we can observe a single, well resolved crystal structure in the case of $d_z=17$~{\AA}, while in the case of $d_z=27$~{\AA} some defects are present. It can also be observed that outside the gap, the IL remains in a disordered, liquid state.

Further clarification can be attained by plotting the ion density distribution profiles inside and outside the gap in Figs.~\ref{fig:cryst_layers}(A) and (B). It can be observed that at the plate-to-plate distance $d_z=17$~{\AA}, both cation and anion peaks of density distribution function inside the gap are narrow and sharp. In addition, both the anion and cation peaks in each paired layer are located at approximately the same $z$ location. These findings confirm that under these conditions the IL is in a crystalline, "solid-like" state with minimum disorder. In the case of a wider gap $d_z=27$~{\AA} the anion peaks next to the walls remain narrow, with a third broader one appearing in the centre. The cation arrangement is more dispersed, with double peaks appearing above and below each anion peak. These statistics indicate a more layered, less strictly ordered state. The difference in the extent of confinement induced crystallisation is a probable reason for the observed steep slope of specific friction since the observed defects can interact more strongly with the upper plate at higher velocities and contribute to the increase of friction force. Our observations show some similarity to the behaviour previously seen in Lennard-Jones systems where systems at pressures above a certain critical value and at sufficiently low velocities exhibited such behaviour. In these studies, cf. Ref.~\cite{Martinie2016}, the shape of fluid molecule was identified as the main parameter that controls crystallisation through the promotion or prevention of internal ordering.

\subsection{Impact of Ionic Liquid pressurisation}

In addition to the impact of different confinement gaps, the effect of IL pressurisation was studied, while the inter--plate distance was kept constant. More specifically, a gap of $d_z=27$~{\AA} was used, while different pressures $p_{\rm ext}=0,$ $120$ and $250$~kPa were applied, using the approach described in Section 2.2.

\begin{figure}[htb]
\includegraphics[width=0.8\textwidth,natwidth=1760pt,natheight=1030pt]{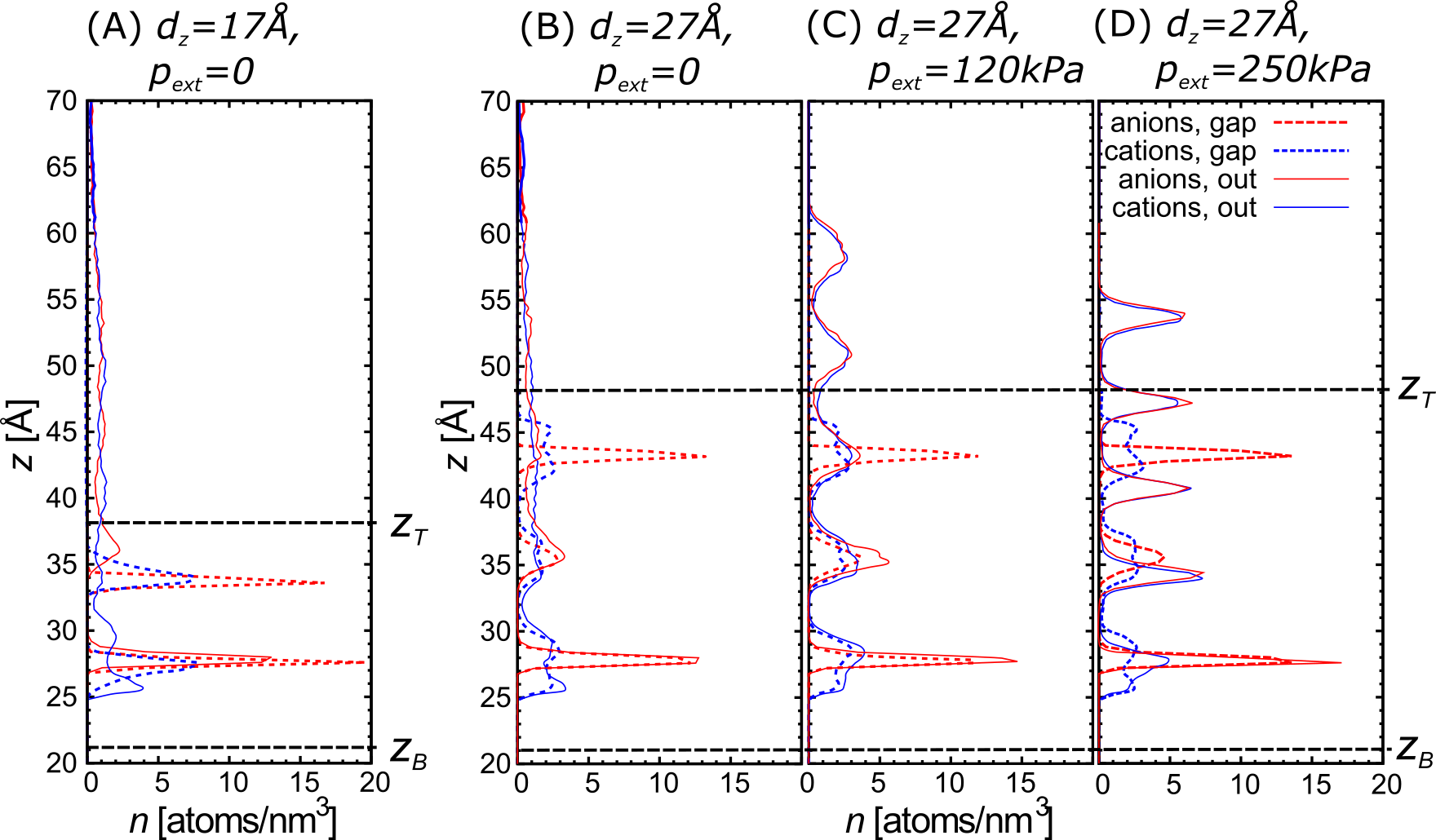}
\caption{Density distributions of ions along the $z$-axis inside (dashed lines) and outside (solid lines) the confinement zone between the solid plates for configurations shown in
Fig.~\ref{fig:cryst_snaps}. The position of the atomic centres of the innermost layer
of the top and bottom plate is denoted with $z_{\rm T}$ and $z_{\rm B}$, respectively. Bottom plate is fixed with $z_{\rm B}=21$~{\AA}.}
\label{fig:cryst_layers}
\end{figure}

Through observation of Fig.~\ref{fig:cryst_layers}(B)-(D), it can be seen that the application of external pressure prevents the wetting of the side walls of the top plate and leads to a distinct crystallisation of the unconfined IL. On the other hand, the ion density profiles inside the confinement zone are moderately influenced.

The friction results for increasing values of applied pressure are consistent with the observations in the previous subsection, with specific friction decreasing as the order of the IL increases. It can be seen that for high external pressure, i.e., $p_{\rm ext}=250$~kPa, the slope of the specific friction curve almost vanishes.

Figure~\ref{fig:cryst_snaps}(C) shows that for $p_{\rm ext}=120$~kPa the local cubic structure induced by confinement between the plates served as a nucleus for further crystallisation between the plates and a well ordered single crystallite was formed in this region. Outside the confinement zone another crystallite was formed with a different orientation. Further increase of external pressure to $p_{\rm ext}=250$~kPa forced the IL in the void space to crystallise, while at the same time the IL in the confinement zone was converted to a number of smaller crystallites,
cf. Fig.~\ref{fig:cryst_snaps}(D) and~\ref{fig:cryst_layers}(D).

The reported results show a dual nature of IL lubrication, with EHL characteristics at low to medium pressures and confinement gaps that allow more than two distinct anion/cation pair layers to form. At higher pressures and smaller distances which can be translated as mixed lubrication conditions the IL is transformed to a solid-like body, while specific friction decreases to low values which are independent of the sliding velocity. This behaviour can be beneficial in engineering applications such as the piston ring--cylinder liner system, where it can be assumed that the IL crystallisation can potentially aid in preventing the solid contact between the surfaces, along with the associated high friction and wear.

\section{Conclusions}

In the current study we have implemented a MD simulation setup in order to study the behaviour of model ionic liquids
confined between plates which are in close proximity while being in relative motion. Our MD setup was developed in a way that allows the meso--scopic study of the lubrication processes in automotive applications such as the piston ring -- cylinder liner interaction inside the internal combustion engine. More specifically, our geometry was selected in order to allow a variable
lubricant confinement gap combined with a varying lubricant quantity in
the gap, while avoiding the squeeze-out of the lubricant into vacuum.
Odd-number layering and near-wall solidification was observed between the solid plates, similar to
published experimental findings. Our friction simulations have uncovered an interesting behaviour of ILs, with a logarithmic dependence of specific friction on velocity hinting at elasto-hydrodynamic lubrication at low loads. This behaviour completely changed under more critical conditions of high load, with specific friction decreasing to lower values and becoming independent of sliding velocity. This behaviour was strongly correlated with the internal structure of the IL and can provide guidance for implementing lubrication concepts that can lead to friction reduction in internal combustion engines.

\section*{Acknowledgments}

The work of I.S. and M.D. was supported in part by the Serbian Ministry of Education, Science and Technological Development under Project No. OI171017, and by COST Action MP1303. Numerical simulations were run on both the HPC facilities of the Advanced Technology Division in Toyota Motor Europe and the PARADOX supercomputing facility at the Scientific Computing Laboratory of the Institute of Physics Belgrade.

\section*{References}

\bibliography{references}

\newpage
\section*{Figures}

\end{document}